\documentclass[a4paper]{jpconf}
\usepackage{mathptmx}
\usepackage{graphicx}
\usepackage{amssymb,amsmath,caption}
\usepackage{mathtools}
\usepackage{tablefootnote}

\begin{document}

\title{Study of neutron shielding collimators for curved beamlines at the European Spallation Source}

\author{V. Santoro $^{1,2}$, D.~D.~DiJulio$^{1,2}$, S.~Ansell$^1$,  N.~Cherkashyna$^1$, G.~Muhrer$^1$  and P.~M.~Bentley$^{1,3}$}

\address{$^1$ European Spallation Source ERIC, 22100 Lund, Sweden}
\address{$^2$ Department of Physics, Lund University, 22100 Lund, Sweden} 
\address{$^3$ Department of Physics and Astronomy, Uppsala University, 75105 Uppsala, Sweden} 

\ead{valentina.santoro@esss.se}

\begin{abstract}
The European Spallation Source is being constructed in Lund, Sweden and is planned to be the world's brightest pulsed spallation neutron source for cold and thermal neutron beams ($\le$ 1 eV). The facility uses a 2 GeV proton beam to produce neutrons from a tungsten target. The neutrons are then moderated in a moderator assembly consisting of both liquid hydrogen and water compartments.
Surrounding the moderator are 22 beamports, which view the moderator's outside surfaces. The beamports are connected to long neutron guides that transport the moderated neutrons to the sample position via reflections. 
As well as the desired moderated neutrons, fast neutrons coming directly from the target can find their way down the beamlines.  These can create unwanted sources of background for the instruments.
To mitigate such a kind of background, several instruments will use curved guides to lose direct line-of-sight (LoS) to the moderator and the target. In addition instruments can also use 
shielding collimators to reduce the amount of fast neutrons further traveling down the guide due to albedo reflections or streaming. Several different materials have been proposed for this purpose.  We present the results of a study of different options for collimators and identify the optimal choices that balance cost, background and activation levels.
\end{abstract}

\begin{figure}[h]
\begin{minipage}{28pc}
\includegraphics[width=28pc]{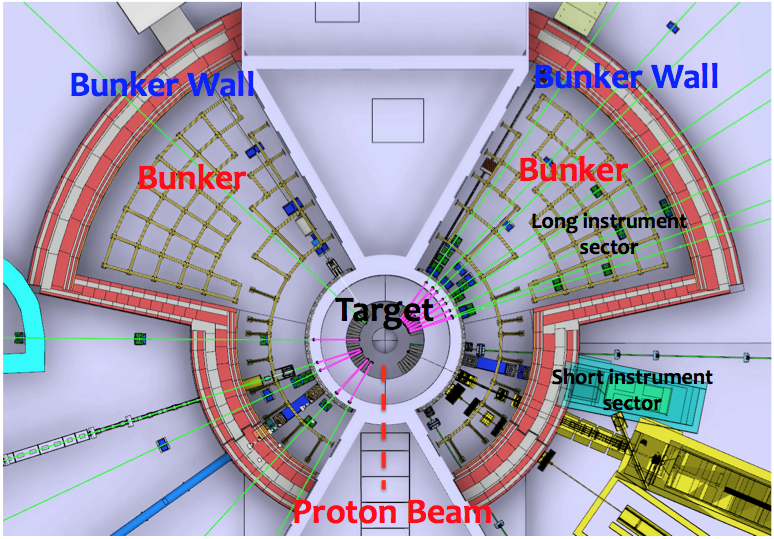}
\caption{\label{fig1}The engineering model of the ESS bunker. }
\end{minipage}\hspace{2pc}\\
\begin{minipage}{30pc}
\includegraphics[width=30pc]{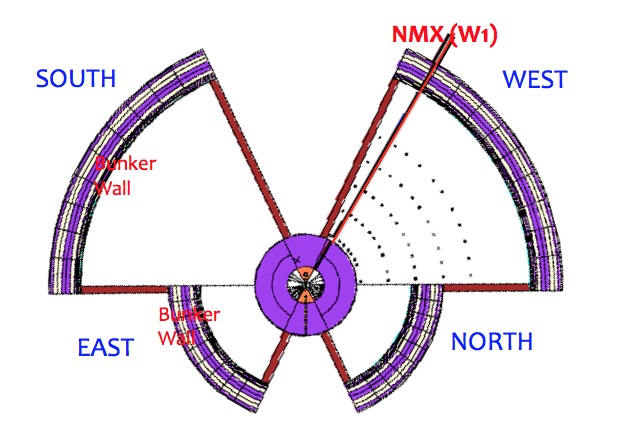}
\caption{\label{fig2} The MCNP6 model of the ESS bunker. All the sectors have been included, the NMX beamline is modeled on the first beam line on the west sector (W1).   }
\end{minipage} 
\end{figure}

\section{Introduction}
\hspace{1cm} \\
\indent The European Spallation Source (ESS)  will be the most powerful neutron source in the world for neutron scattering
studies of condensed matter. Neutrons will be produced by a 2 GeV proton beam impinging on a target made of tungsten.
The neutrons will be eventually  slowed down by the moderator placed next to the target. A reflector
will surround the moderator to increase the neutron intensity.
The tungsten target, moderator and reflector will be located inside the target monolith that will be shielded by a 3.5~m of steel which extends to a radius of 5.5~m from the moderator center.
Within this monolith shield wall there will be beam extraction ports for each instrument. The monolith wall will shield the target and
reduce the radiation escaping from it. Nevertheless, the neutron radiation dose coming out from the monolith is still substantial and must be shielded and contained in an additional  common shielding structure called the bunker. \\
\indent  The ESS bunker will extend from the outer surface of the monolith located at 5.5~m from the target to 15~m for the short instrument (instruments several tens of meters )
and to 28~m for long instruments (instruments up to $\sim$150~m in length). The current design of the ESS bunker is essentially a contained void with a 1.7 m thick roof about 3 m above the internal bunker floor and a 3.5 m thick wall, starting at 11.5 m for the short instrument sector or 24.5 m for the long instrument sector, as can be seen in figure~\ref{fig1}.
The open area inside the bunker will be filled with components related to the instruments such as metallic guides, choppers, shutters or collimators.
These components will receive a substantial radiation dose during beam on target and they will be activated.
Rapid access to the equipment in this area is required for performing maintenance and repairs.  For this reason, the optimisation of the activation level plays an important role in determining the serviceability and time-scales for
accessing these components.
In this paper we focus on the neutron shielding collimators for curved beam lines.
The choice of material for such a component is a compromise between the efficiency of reducing the neutron radiation dose, the cost, and the activation of the material.

\begin{table}
\caption{\label{parameters}NMX beamline parameters as used in the simulation. }
\begin{center}

\includegraphics[width=30pc]{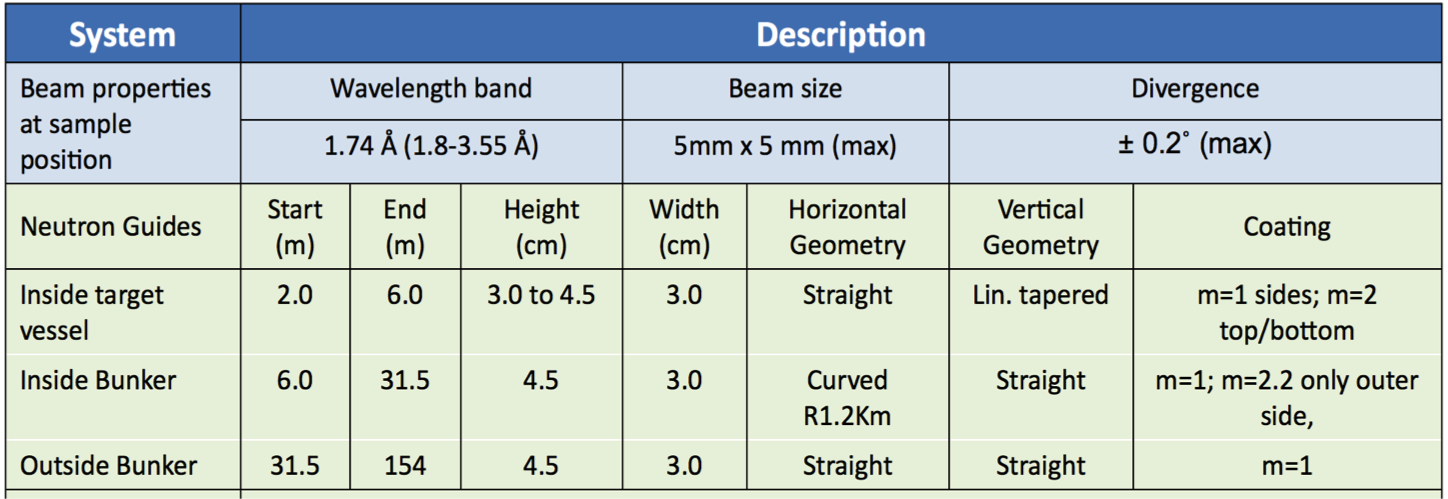}

\end{center}
\end{table}

\begin{figure}[htp]
\begin{minipage}{28pc}
\includegraphics[width=30pc]{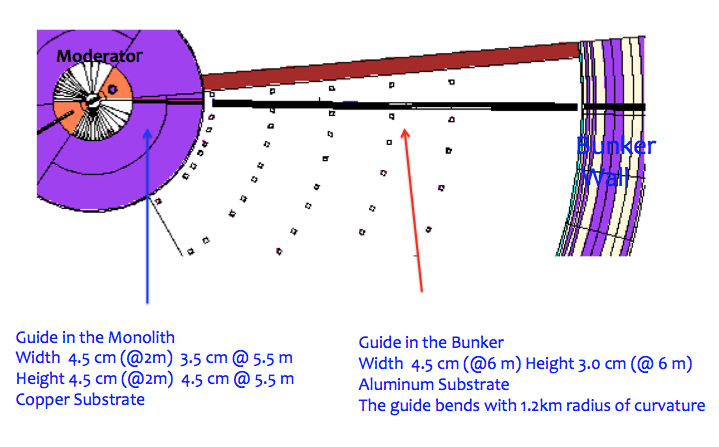}
\caption{\label{geo1}Overview of the NMX beamline in the bunker .}
\end{minipage}\hspace{2pc}\\
\begin{minipage}{28pc}
\includegraphics[width=32pc]{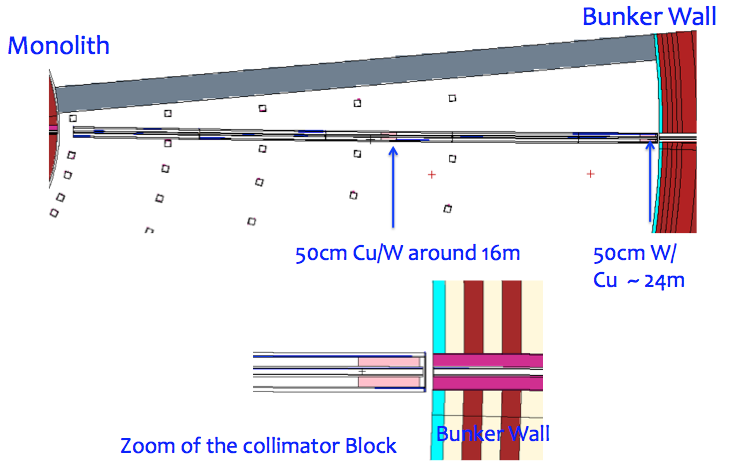}
\caption{\label{geo2}Collimator blocks in the bunkers. In the current picture one collimator is located at 16~m from the moderator and the other one is just before the bunker wall (24~m from the moderator).}
\end{minipage} 
\end{figure}

\section{The NMX instrument}
\hspace{1cm} \\
\indent For the collimator optimisation study we will use the NMX~\cite{nmx} instrument that is a time-of-flight (TOF)
quasi-Laue diffractometer optimised for small samples and large unit cells. A table with the  beamline parameters in its original design 
and used in this simulation can be found in table~\ref{parameters}.
The NMX instrument is basically composed of a straight  guide in the monolith and a  bender in the bunker. 
The curving of the guide permits the loss of line of sight to the moderator and target and the use of collimators allows the stripping out of the fast neutron component traveling down the guide. \\
\indent To perform these studies we used the radiation transport code MCNP6~\cite{mcnp}. 
The MCNP6 model has been built using Comblayer~\cite{comb} which is a C++ package that can write MCNP input files with a detailed geometrical description.
A snapshot of the MCNP6 model for the ESS created with Comblayer is shown in figure~\ref{fig2}. 
As can be seen from the picture the model contains several details like the exact geometry of the target wheel, the moderator and the monolith and is a very close representation of the current engineering model (figure~\ref{fig1}). \\
\indent The NMX beamline is located in the long instrument west sector and since it is the first beamline of this sector it is identified with the beamport W1. 
The beamline has been modelled in details with the guide in the monolith, the vacuum pipe, the vacuum window and the guide in the bunker,  with the addition of two collimators. Figure~\ref{geo1} shows an overview of the beamline, while figure~\ref{geo2} shows the possible locations for the collimators in the bunker. 
As stated above, the primary objective of this study is to design the collimator blocks to minimise the fast neutron component that is travelling inside the guide. Reducing this radiation is not only advantageous for background reasons but can also considerably reduce the shielding cost further downstream of the collimator position.

\section{Optimisation of the collimators blocks }
\begin{figure}[h]
\begin{minipage}{30pc}
\includegraphics[width=30pc]{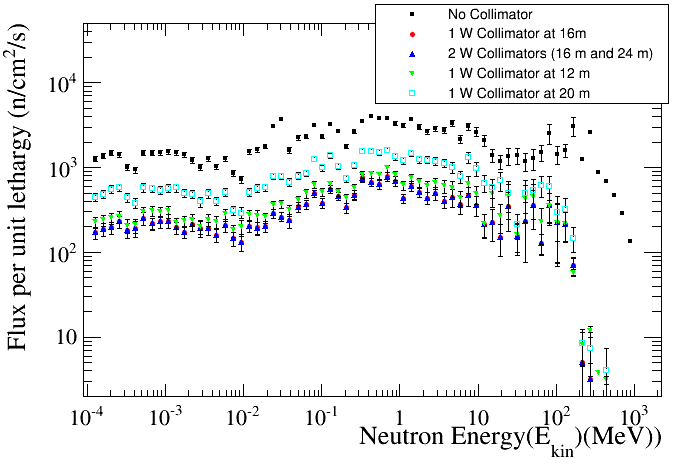}
\caption{\label{spec1}Fast neutron energy spectrum for different collimator configurations. The difference between the neutron energy spectrum from one tungsten collimator at 16 m (red dots) and the use of an additional collimator at 24 m (blue dots) could not be seen since they completely overlap. The reason is due to the fact that the bunker insert is sufficient to remove neutrons that have not stripped off by the first collimator. }
\end{minipage}\hspace{2pc}%
\end{figure}

\begin{figure}[h]
\begin{minipage}{30pc}
\includegraphics[width=30pc]{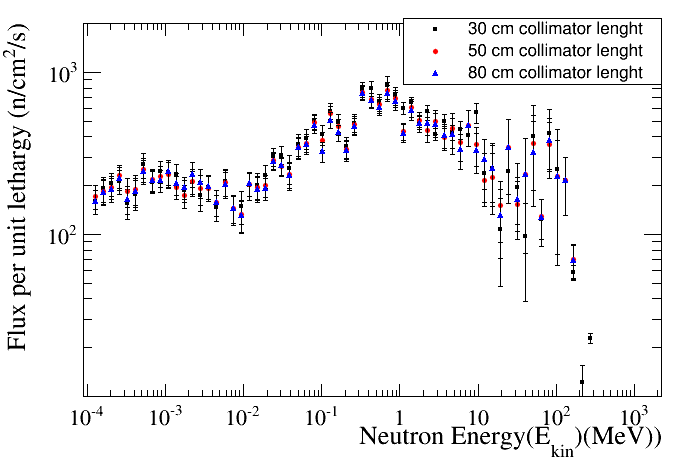}
\caption{\label{spec2}Fast neutron energy spectrum for different collimator lengths.}
\end{minipage}\hspace{2pc}%
\end{figure}

\begin{figure}[h]
\begin{minipage}{30pc}
\includegraphics[width=30pc]{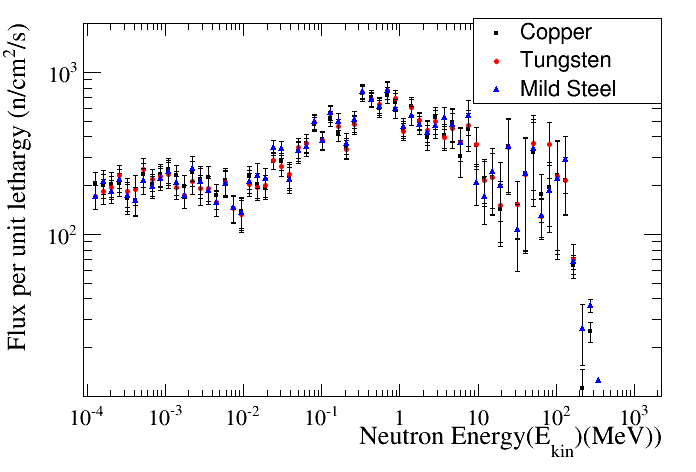}
\caption{\label{spec3}Fast neutron energy spectrum for different collimator materials. The tungsten and the copper used correspond to pure material while the steel composition is given in table~\ref{mytable}.}
\end{minipage}\hspace{2pc}%
\end{figure}

\hspace{1cm} \\
\indent To study what is the best configuration that allows a reduction of the fast neutron component we calculate the neutron energy spectrum immediately after the bunker wall (28~m after the moderator center) in the guide, for different collimators options and positions. 
The results of these calculations can be seen in figure~\ref{spec1},~\ref{spec2}~and~\ref{spec3}. 
The use of collimators, as can be seen from the spectra shown in figure~\ref{spec1}, significantly reduces the high energy component but there is also a reduction  in the keV energy region of the spectrum. Among all the possibilities studied, the best option is to use one collimator around 16~m where the guide started to lose significantly line of sight.
Locating a collimator upper stream (12~m from the moderator) or downstream  (20~m from the moderator) does not have the same effect.
Adding another collimator just before the bunker wall has no significant effect (from figure~\ref{spec1} the difference between the spectrum for one collimator at 16~m (red dots) and the one with the addition of a second collimator at 24~m (blue dots) can not be seen) since the bunker insert is sufficient to remove neutrons that have not been stripped off by the first collimator. 
In all these studies, the length of the collimator was 50~cm and the material used was pure tungsten.
An additional calculation to establish what is the effect of decreasing or increasing the length of the collimator is shown in figure~\ref{spec2}. The results show that a 30~cm collimator provides the same performance as a 50~cm or 80~cm collimator. Thus increasing the length beyond 30~cm
adds no performance. \\
\indent Once we found the best position and the ideal length for the collimator, we compared different materials. The plot in figure~\ref{spec3} shows a comparison between pure copper, pure tungsten and mild steel (mild steel is  a low carbon steel, its composition is shown in table~\ref{spec3}).  As can be seen in the figure, the performance of the three materials is the same.

\section{Activation  of the collimators block}
\begin{figure}[h]
\begin{center}
\includegraphics[width=15pc]{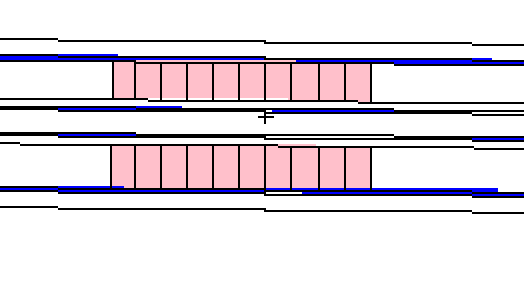}
\caption{\label{collimatorsection} Side view of the collimator. The collimator has been divided into several different segments .}
\end{center}
\end{figure}

\hspace{1cm} \\
\indent We observed that the fast neutron reduction for the collimators are very similar for different materials.
To find the best option however it is necessary to study their activation levels after the beam is off, since as explained before the collimators will be located in the ESS bunker area where maintenance and repairs of different components will be carried out. All the activation calculations have been performed using CINDER 1.05~\cite{cinder}.
For this calculation we used pure tungsten and pure copper, while for the steel we used a low carbon steel. The composition of the steel can be found in table~\ref{mytable}.
The activation levels were studied using the collimator located at 16 m and by dividing it into several different segments (see figure~\ref{collimatorsection}) since the first part of the collimator will receive a higher flux of neutrons and become more active. 

\begin{figure}[h]
\begin{minipage}{35pc}
\includegraphics[width=35pc]{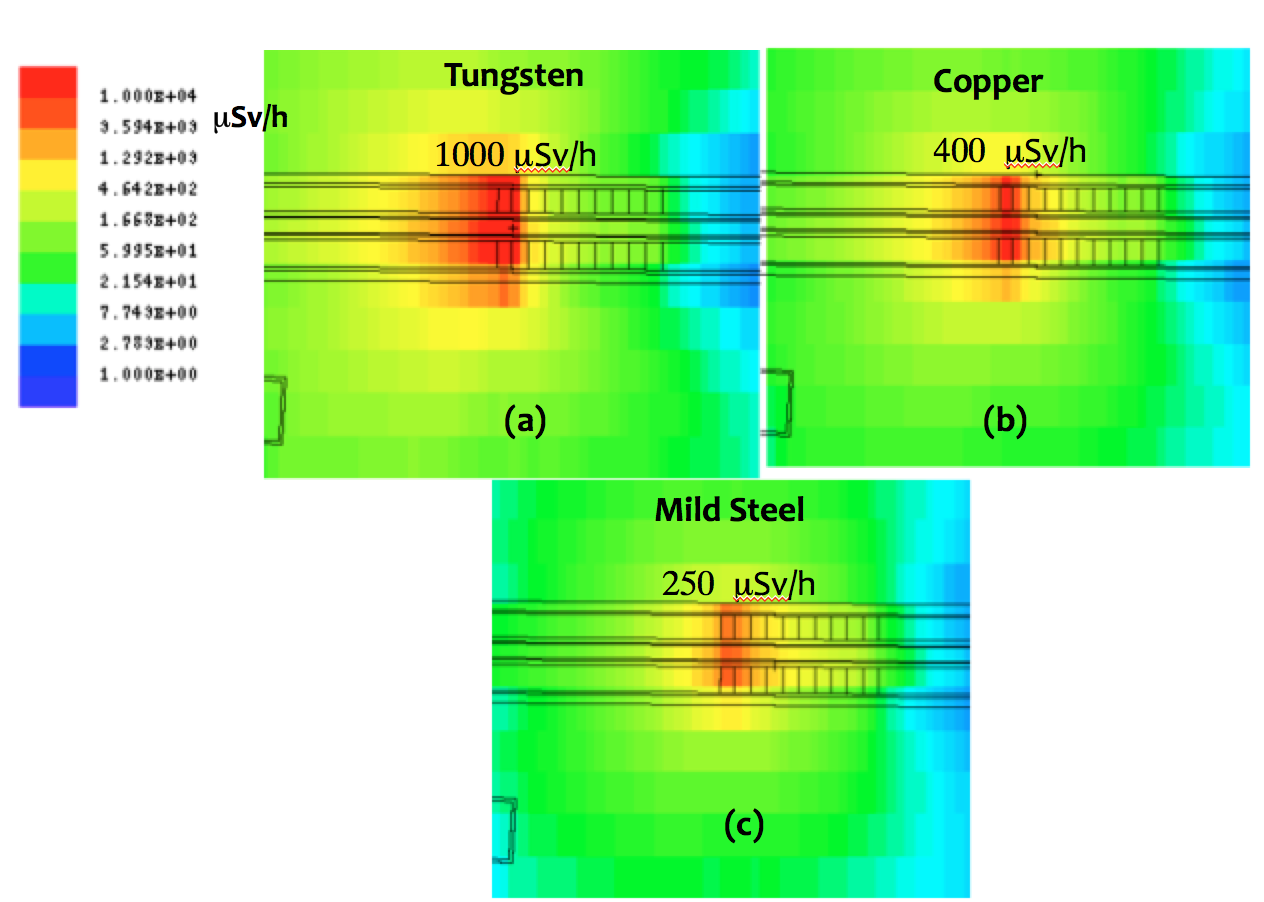}
\caption{\label{act1} Photon radiation dose rates for the NMX collimator block located around 16~m from the moderator after one day of beam off. (a) Tungsten (b) Copper and (c) Mild Steel (see table~\ref{mytable}).}
\end{minipage}\hspace{2pc}%
\end{figure}

\begin{figure}[h]
\begin{minipage}{35pc}
\includegraphics[width=35pc]{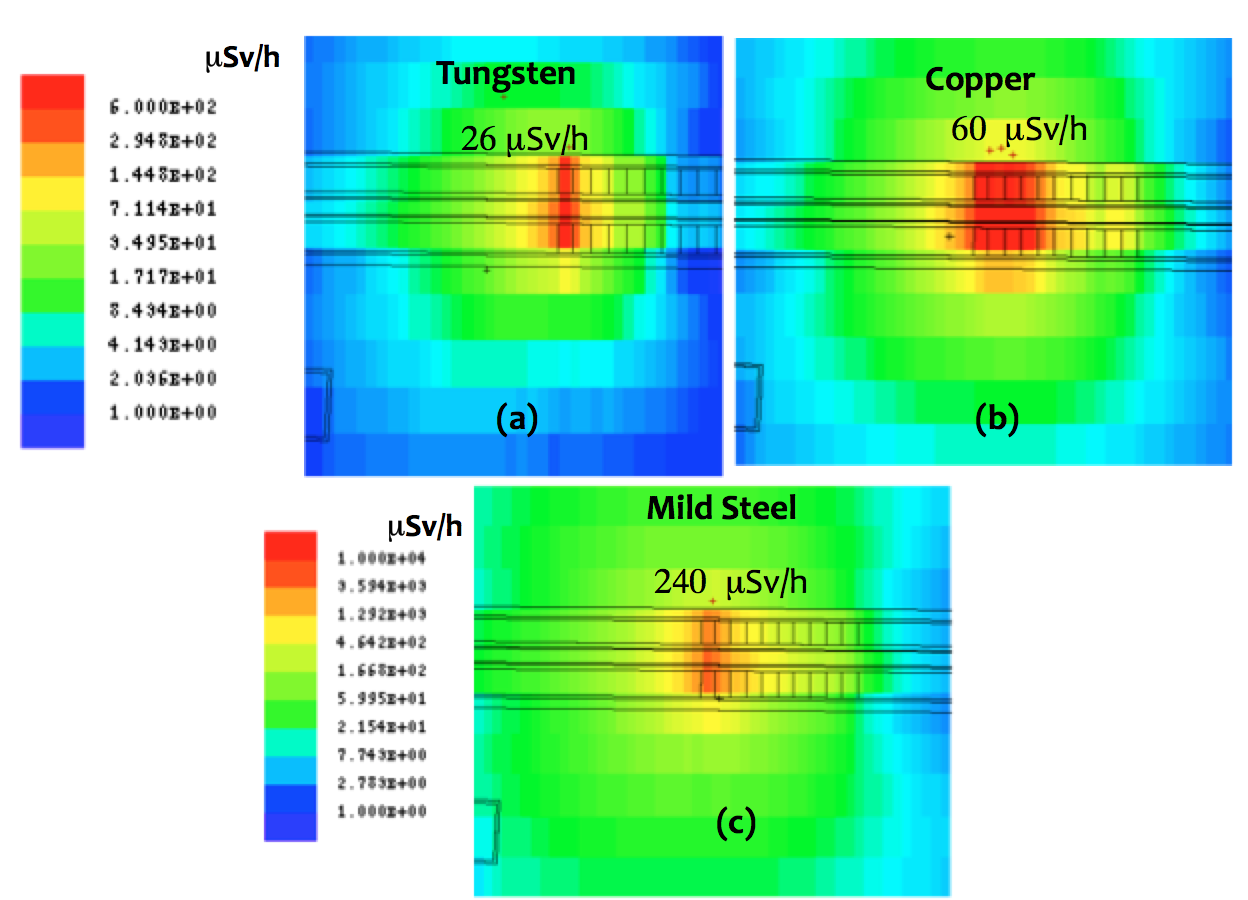}
\caption{\label{act2}Photon radiation dose rates for the NMX collimator block located around 16~m from the moderator after seven days of beam off. (a) Tungsten (b) Copper (c) and Mild Steel (see table~\ref{mytable}).}
\end{minipage}\hspace{2pc}%
\end{figure}

The activation of the collimator block is computed after 10 years of irradiation with the proton beam energy of 2~GeV at the accelerator power of 5~MW.
Figure~\ref{act1} shows the activation for the different materials after 1 day of beam off. The numbers shown in the figures represent the photon dose at 5 cm away from  the highest hot spot of the collimator. Figure ~\ref{act2} shows the activation level after 7 days of beam off.
As can be seen from the plots the activity of the collimator is not uniform across its length. The tungsten and the copper block will get quite active after one day
but they will decay to a quite lower value after one week, while the mild steel will be still be very active. The reason for the different behaviour of the materials is due to 
the different half lives that drive the activation of the collimator.
The tungsten activation after one day of beam off is mainly due to decays of $^{187}$W (this isotopes has a half life of 23.72 hours), that will be mostly gone after one week, at that point the main decays will be due to $^{175}$Hf and $^{182}$Ta.
The copper block activation after one day is driven by the decays of $^{64}$Cu (half life of 12.7 hours) while 
after one week is mainly due to decays of $^{60}$Co. 
For the mild steel the activation level is quite similar from 1 day to seven days since it is driven by the $^{60}$Co decay that has a half life of more than five years.

\begin{table}
\caption{\label{mytable}Mild Steel composition used in the simulations .}
\begin{center}
\begin{tabular}{llll}
\br
Element  & Atomic Mass  & atom (\%) \\
\mr
 Mild Steel & &\\
~~~~~B &~~~~~10 & 3.99~$\times$~10$^{-5}$ & \\
~~~~~C &~~~~~12 & 1.20~$\times$~10$^{-3}$ & \\
~~~~~N &~~~~~14 & 8.00~$\times$~10$^{-5}$ & \\
~~~~~Al &~~~~~27 & 1.27~$\times$~10$^{-3}$ & \\
~~~~~Si &~~~~~28 & 9.96~$\times$~10$^{-4}$ & \\
~~~~~P &~~~~~31 & 5.00~$\times$~10$^{-4}$ & \\
~~~~~S &~~~~~32 & 4.00~$\times$~10$^{-4}$ & \\
~~~~~Ti &~~~~~47 & 1.00~$\times$~10$^{-4}$ & \\
~~~~~V &~~~~~51 & 1.00~$\times$~10$^{-4}$  & \\
~~~~~Cr &~~~~~52 & 1.14~$\times$~10$^{-3}$  & \\
~~~~~Fe &~~~~~54 & ~~0.986  & \\
~~~~~Mn &~~~~~55 & 4.00~$\times$~10$^{-3}$ & \\
~~~~~Ni &~~~~~58 & 1.00~$\times$~10$^{-3}$ & \\
~~~~~Co &~~~~~59 & 2.00~$\times$~10$^{-4}$ & \\
~~~~~Cu &~~~~~63 & 1.50~$\times$~10$^{-3}$ & \\
~~~~~Mo &~~~~~92 & 1.00~$\times$~10$^{-3}$ & \\
~~~~~Nb &~~~~~93 & 1.00~$\times$~10$^{-4}$ & \\
~~~~~Sn &~~~~~118 & 9.93~$\times$~10$^{-5}$ & \\
\br
\end{tabular}
\end{center}
\end{table}

\section{Conclusions}
\hspace{1cm} \\

Based on these simulations, it is recommended to use at most 30~cm of copper or tungsten at 16m around the vacuum pipe for NMX.
It is important to underline that for the copper and tungsten in the simulations pure materials have been assumed. If instead of the pure material 
an alloy will be used, the presence of impurities could change significantly the activation results.


\section*{References}
\hspace{1cm} \\


\begin{thebibliography}{99}
\bibitem{nmx}
https://europeanspallationsource.se/instruments/nmx 
\bibitem{mcnp}
T. Goorley, et. al., "Features of MCNP6", Supercomputing in Nuclear Applications and Monte Carlo 2013, Paris, Oct 27-31, LA-UR-13-28114 (2013).
\bibitem{comb} 
S. Ansell, "CombLayer: A fast parametric MCNP(X) model constructor",
Proceedings of the 21st Meeting of the International Collaboration on
Advanced Neutron Sources, Mito, Japan, Feb. 2016.
\bibitem{cinder}
W. B. Wilson, S. T. Cowell, T. R. England, A. C. Hayes \& P. Moller:
A Manual for CINDER'90 Version 07.4 Codes and Data, LA-UR-07-8412 


\end{thebibliography}
\end{document}